---------------------------------------------------------------

\documentstyle[prl,aps]{revtex}
\begin{document}
\preprint{UG-DFM-11/95}
\title{ABSENCE OF CLASSICAL AND QUANTUM MIXING}
\author{L.L. Salcedo}
\address{Departamento de F\'{\i}sica Moderna, Universidad de Granada,
E-18071 Granada, Spain }
\maketitle
\begin{abstract}
It is shown, under mild assumptions, that classical degrees of freedom
dynamically coupled to quantum ones do not inherit their quantum fluctuations.
It is further shown that, if the assumptions are strengthen by imposing
the existence of a canonical structure,
only purely classical or purely quantum dynamics are allowed.
\end{abstract}
\pacs{03.65.Bz}

1. There is consensus among physicists that Quantum Mechanics is the
correct description of Nature, at least within the range of presently
observable scales. Nevertheless, some systems are routinely
described using a classical, and thus approximated, dynamics. This can
be so either for simplicity or due to the lack of a consistent quantum
theory. Einstein's General Relativity is an example of the latter.
In an excellent speculative paper \cite{BT88}, Boucher and Traschen
consider several physical systems which require
a mixed description in terms of
quantum and classical degrees of freedom, mutually interacting.
A good example is provided by early universe physics, where
fully quantum matter fields are coupled to classical gravitational fields.
The traditional approach to this problem has been to couple the
gravitational fields to the expectation values of the quantum
energy-momentum tensor, see e.g. Ref. \cite{Br85}.
This kind of approach has been criticized \cite{BT88},
on the grounds that the classical fields
evolve deterministically, hence,
the quantum fluctuations in this fields,
induced by their coupling to the quantum fields, are missed.

This criticism (as well as presumably the challenge it presents)
has led to look for a mathematically consistent
description of semiquantized systems, i.e., mixed classical-quantum
systems \cite{Al81,BT88,An95,Sa94,Di95a}.
These systems are considered by themselves, that is,
not as the limit of a fully quantum theory.
The fact that the classical description is
just an approximation is disregarded in this context,
since the purpose is to define a
mathematical structure with some physical input.

In this letter it is shown that in fact there are severe obstructions to
construct such a description and, if it exists at all, it will not enjoy
the elegant mathematical structures common to classical or
quantum mechanics.
Since presently there is no widely accepted definition of what is meant by a
semiquantized system, and in order not to discard potentially
interesting choices, we should rely on properties as general as
possible, which must hold, in particular, for the purely classical and
purely quantum cases. Throughout, the degrees of freedom will be bosonic,
though this assumption does not seem to be essential for the arguments.

2. It is assumed (i) that the set of observables
forms an associative algebra ${\cal A}$ over the field of complex numbers.
Let ${\cal A}_L$ be the algebra spanned by the coordinates $q_i$, the
conjugate momenta $p_i$, $i=1,\dots,N$ and the identity $E$, as
generators, i.e., the set of formal series of ordered products of them.
Then the physical algebra ${\cal A}$ is defined as the quotient algebra of
${\cal A}_L$ modulo some identities among the generators
(e.g., commutations relations). These identities characterize the
algebra and are to be specified.
${\cal A}$ will be a non-commutative algebra in general.
By definition, $A E=EA=A$ for every observable $A$.
As a consequence, $E$ is the only element with this property.
In classical mechanics ${\cal A}$ is just the set
of complex functions in phase space and $E$ is the unity function.
In quantum mechanics we have the algebra of operators in the Hilbert
space of the system. Here, the word observable is being used in a slightly
wider sense that usual, since it includes non-real functions and
non-Hermitian operators as well. This axiom is also present in
Refs. \cite{BT88,An95},

A second axiom (ii) refers to the time evolution of the observables
(Heisenberg picture). Namely, there is a family ${\cal U}$ of evolution
operators ${\cal U}(t_1,t_2)$ in ${\cal A}$
such that ${\cal U}(t_1,t_1)$ is the identity operator and
${\cal U}(t_1,t_2){\cal U}(t_2,t_3)={\cal U}(t_1,t_3)$.
Furthermore, the evolution preserves the algebraic structure, that is,
if two observables $A_{1,2}(t_0)$ evolve to $A_{1,2}(t)$, and $c_{1,2}$ are
constant complex numbers, $c_1 A_1(t_0)+c_2 A_2(t_0)$ evolves to
$c_1 A_1(t)+c_2 A_2(t)$, and $A_1(t_0)A_2(t_0)$ evolves to $A_1(t)A_2(t)$.
In other words, time evolution forms a grupoid of algebra
automorphisms of ${\cal A}$. Certainly, this axiom holds both in
classical and in quantum mechanics, and it is hard to imagine an
interesting formulation which would violate it. Moreover, the
endomorphism property follows
from the Schr\"odinger picture, since there the observables do not
evolve, and hence the algebraic structure is trivially preserved.
Note that ${\cal U}$ gives only the dynamic time dependence of
observables, and also that, in principle, the operator ${\cal U}(t_1,t_2)$
does not correspond to an algebra element (e.g., in the
purely classical case).
On the other hand, it is not assumed that the system is conservative.
There can be time dependent external fields which break invariance
under time translations. Similarly, time reversal invariance is not required.

Some relevant conclusions can be extracted from these two axioms.
If a set of elements generates the algebra, this property is
maintained through time evolution. This follows from time evolution
being an automorphism.
The observable $E$ is time independent: for any observable $A$,
the relation $EA=A$ evolves to $E(t)A(t)=A(t)$ and thus
$E(t)=E$, using that $A(t)$ is an arbitrary observable, since time
evolution is a bijection and $A$ was arbitrary. Another consequence is that
commutation relations of the form $[A(t_0),B(t_0)]=cE$
are also preserved, since
they evolve to $[A(t),B(t)]=cE$. In particular, if two
observables commute at any given time they do so at any time.

A last axiom (iii) is needed referring to the commutation relations.
The classical
dynamics is characterized by commuting coordinates and momenta which
evolve according to Hamilton's equations. On the other hand, the
quantum dynamics satisfies the canonical commutation relations and the
Heisenberg evolution equation, $dA/dt=i[H(t),A]$.
For the semiquantum dynamics it is postulated that
the classical commutation relations hold among the classical
generators and similarly for the quantum sector.
Furthermore, the generators of the classical sector commute with those
of the quantum one.
In other words, the commutation relations are as follows:
\begin{equation}
[q_i,q_j]=[p_i,p_j]=0\,,\qquad
[q_i,p_j]=i\lambda_i\delta_{ij}E\,,\qquad i,j=1,\dots,N\,,
\label{eq1}
\end{equation}
where $\lambda_i$ is zero if $i$ is the label of a classical degree of
freedom, and unity (or $\hbar$) if it labels a quantum one.
These are the defining identities of the algebra of the semiquantized system.

This axiom can be justified as follows. Certainly, eqs.~(\ref{eq1})
are natural if the semiquantized system consists of a
classical sector and a quantum sector without any interaction
among them; this is a physical assumption. Since both in classical and
quantum dynamics the commutation relations are unaffected by the
choice of the interaction, one should expect that this is true as
well in the semiquantized case, and hence eqs.~(\ref{eq1}) follow.
Another argument can be given by introducing a second physical
assumption, namely, that the coupling among the two sectors can be switched
on and off by playing with suitable time dependent coupling constants.
Now, we can imagine starting with an uncoupled system, which
satisfies the relations (\ref{eq1}), switching on the interaction to
end up with any given fully coupled system. Since the commutation
relations are preserved by time evolution (even for non conservative dynamics)
eqs.~(\ref{eq1}) will hold too in an arbitrary coupled semiquantized system.
We think that these considerations make axiom (iii) inescapable.

Now, from the previous considerations, a quite strong result
can be derived, namely, the subalgebra spanned by the classical sector is
invariant under time evolution. To simplify the notation,
let us consider a system with just two degrees of freedom, one of
them, $(q_1,p_1)$, quantum and the other $(q_2,p_2)$ classical, i.e.,
$\lambda_1=1$ and $\lambda_2=0$ in eqs.~(\ref{eq1}). Further,
the coordinates and momenta at $t=t_0$ are denoted by $q_i$, and
$p_i$, respectively.
For any time $t$, the set $\{E,q_1(t),p_1(t),q_2(t),p_2(t)\}$
generates the whole algebra ${\cal A}$, and $q_2(t)$ commutes with all
of them from eqs.~(\ref{eq1}), thus $q_2(t)$ commutes with all the algebra
elements and, in particular, with $q_1$ and $p_1$, and the same
holds for $p_2(t)$. On the other hand, again using the commutation
relations, every $A\in {\cal A}$ is uniquely characterized by a
set of coefficients $c_{k\ell mn}$, $k,\ell,m,n=0,1,\dots$, as
$A=\sum_{k\ell mn}c_{k\ell mn}q_1^kp_1^\ell q_2^mp_2^nE$.
It is immediate to see that any element commuting with $q_1$ cannot
contain $p_1$ and vice versa. Therefore, $q_2(t)$
must be of the form
$\sum_{mn}c_{mn}(t)q_2^mp_2^nE$, and similarly $p_2(t)$. In
other words, $q_2$ and $p_2$ are commuting objects which
evolve by themselves following classical trajectories, without fluctuations.
On the other hand,
$q_1(t)$ and $p_1(t)$ may depend on $q_2(t)$ and $p_2(t)$ which,
in this regard, behave as external sources.

One realization of the above picture is the traditional approach to
semiquantization, namely, the quantum degrees of freedom move in the
presence of the classical background. On the other hand, the
classical degrees of freedom are coupled to the expectation values of
the quantum variables.
For instance, if the system consists of two coupled
harmonic oscillators, the equations of motion take the following form:
\begin{mathletters}
\label{eq2}
\begin{equation}
\frac{dq_1(t)}{dt} = \frac{p_1(t)}{m_1}\,,\qquad \frac{dp_1(t)}{dt} =
-m_1\omega_1^2q_1(t) - g(t)q_2(t) \,,
\label{eq2a}
\end{equation}
\begin{equation}
\frac{dq_2(t)}{dt} = \frac{p_2(t)}{m_2}\,,\qquad \frac{dp_2(t)}{dt} =
-m_2\omega_2^2q_2(t) - g(t)\langle q_1(t)\rangle_{\psi_1} \,.
\label{eq2b}
\end{equation}
\end{mathletters}
Here, $\psi_1$ is the state of the quantum sector in Heisenberg picture, i.e.,
a certain time independent wavefunction in the Hilbert space
of $q_1$ and $p_1$. Technically, our axioms apply here by considering
$\langle q_1\rangle_{\psi_1}$ and $\langle p_1\rangle_{\psi_1}$ as fixed
parameters, that is, independent of $q_{1,2}$ and $p_{1,2}$.
Indeed, we can take expectation values of eqs.~(\ref{eq2a}) in
$\psi_1$, and solve the resulting system for $q_2(t)$ and $p_2(t)$;
they will depend dynamically on $q_2$, $p_2$ and $t$ (as well as on
the fixed parameters $\langle q_1\rangle_{\psi_1}$ and
$\langle p_1\rangle_{\psi_1}$).
Substituting the classical solution in eqs.~(\ref{eq2a}),
$q_1(t)$ and $p_1(t)$ are obtained as functions of
$q_1$, $p_1$, $q_2$, $p_2$ and $t$.
Afterwards, to extract meaningful physical results, one must choose precisely
$\psi_1$ as the state of the quantum sector in Heisenberg picture, but
this is not required by our axioms.
It is immediate to check that eqs.~(\ref{eq2})
preserve the commutation relations~(\ref{eq1}): $q_2(t)$ and $p_2(t)$ are just
ordinary functions and hence are commuting objects; $q_1(t)$ and
$p_1(t)$ describe a purely quantum harmonic oscillator coupled to an applied
external force $-g(t)q_2(t)$.

If one insisted in keeping the operator
$q_1(t)$ in eq.~(\ref{eq2b}), instead of its expectation value, i.e.,
\begin{equation}
\frac{dp_2(t)}{dt} = -m_2\omega_2^2q_2(t) - g(t) q_1(t)\,,
\label{eq3}
\end{equation}
a violation of the commutation relations would result.
For instance, assuming that eqs.~(\ref{eq1}) hold at $t=0$,
and, there, treating $\lambda_{1,2}$ as free parameters,
one would find
$[p_1(t),p_2(t)]= (\lambda_1-\lambda_2)ig(0)tE + {\cal O}(t^2)$, which
only vanishes if either $g=0$ and thus the two subsystems are decoupled,
or else if $\lambda_1=\lambda_2$, i.e., the purely
classical case if they vanish, or the purely quantum case, if they do not.
Similarly, $[q_1(t),p_2(t)]$ would break down at ${\cal O}(t^2)$.

3. The canonical structure of both classical and quantum
mechanics (Poisson bracket and commutator, respectively)
has been invoked in the literature \cite{BT88,An95}, as a guiding principle
to define semiquantized theories. From this point of view, it is of
interest to consider whether there exist canonical structures interpolating
between the quantum and the classical limits.

Let us then study which new constraints are found if, in addition to
previous assumptions (i-iii), a canonical structure is present.
For convenience, the relations~(\ref{eq1}) are rewritten in the form:
\begin{equation}
[\phi^\alpha,\phi^\beta]=\eta^{\alpha\beta}E\,,\qquad
\alpha,\beta = 1,\dots, 2N \,,
\label{eq4}
\end{equation}
where the single symbol $\phi^\alpha$ has been introduced to
denote both $q_i$ and $p_i$, and $\eta^{\alpha\beta}$
is an antisymmetric tensor with complex components.

The canonical structure is introduced by three new postulates. First,
there exists (iv) a Lie bracket $(\ ,\ )$ in ${\cal A}$,
which generates the (infinitesimal) canonical transformations by
$\delta_AB=(A,B)$, $A,B\in{\cal A}$, and in particular time evolution
is a canonical transformation
\begin{equation}
\frac{dA(t)}{dt} = (H(t),A(t))\,,
\label{eq5}
\end{equation}
where $H(t)\in{\cal A}$ is the Hamiltonian of the system.
Second, it is assumed (v) that the canonical transformations are algebra
automorphisms. Of course, axioms (iv) and (v) imply (ii).
And third, the following canonical relations are assumed (vi):
\begin{equation}
(\phi^\alpha,\phi^\beta)=\epsilon^{\alpha\beta}E\,,\qquad
\alpha,\beta = 1,\dots, 2N\,,
\label{eq6}
\end{equation}
where $\epsilon^{\alpha\beta}$ is the usual simplectic matrix, namely, zero
for $(q,q)$ or $(p,p)$ and $\delta_{ij}$ for $(q_i,p_j)$, common to
classical and quantum mechanics.

Since $(\ ,\ )$ is a Lie bracket, it is bilinear, antisymmetric and
satisfies Jacobi's identity. This is a consistency requirement among
canonical transformations, which guarantees that
$\delta_A(B,C)=(\delta_AB,C)+(B,\delta_AC)$, i.e., the bracket itself
is invariant. In particular,
the relationship $(A,B)(t)=(A(t),B(t))$ will be consistent with the
equations of motion.
The endomorphism property of the canonical transformations implies
that the Lie bracket is a derivation in ${\cal A}$, i.e., it satisfies
the product (Leibniz) rule: $(A,BC)=(A,B)C+B(A,C)$.
{}From here it is immediate to deduce that
$E$ is invariant under canonical transformations, that is,
$\delta_AE=(A,E)=0$.
A consequence of the two previous observations is that the canonical
relations between the $\phi^\alpha$ are preserved by canonical transformations.

As noted in \cite{Di95} (see also \cite{An95a}),
the brackets defined in Refs.~\cite{BT88} or \cite{An95} are not
derivations, thus the algebraic structure among observables is not
preserved under time evolution (thus, violating axiom (ii))
and this seems unphysical.
Also they are not Lie brackets, since they fail to satisfy Jacobi's identity.
As a consequence, the canonical relations are not preserved either.
Actually, the bracket defined in \cite{An95} is not even antisymmetric,
hence, in general, the energy is not conserved
even by time independent Hamiltonians, and hermiticity is broken
by the dynamic evolution \cite{Di95,An95a}.

We still have to determined in which cases the canonical structure is
consistent with axioms (i-iii).
Using only linearity, antisymmetry, the product rule and
the canonical relations~(\ref{eq6}),
the bracket of every two observables can be worked out and $(\ ,\ )$
becomes completely determined.
Hence, it can be checked whether it admits the commutation
relations~(\ref{eq4}).
Indeed, arbitrary canonical transformations of both sides in~(\ref{eq4}) must
coincide. For any $\alpha,\beta,\mu,\nu=1,\dots,2N$, we find:
\begin{eqnarray}
0 &= (\phi^\alpha\phi^\beta,\eta^{\mu\nu}E) =
(\phi^\alpha\phi^\beta,[\phi^\mu,\phi^\nu]) \nonumber \\
 &= \epsilon^{\beta\mu}\eta^{\alpha\nu}+
\epsilon^{\alpha\mu}\eta^{\beta\nu} +
\epsilon^{\beta\nu}\eta^{\mu\alpha}+
\epsilon^{\alpha\nu}\eta^{\mu\beta} \label{eq7}\\
\nonumber
\end{eqnarray}
The last line follows from repeatedly applying the product rule.
Contracting this equation with $\epsilon_{\mu\beta}$, the inverse
matrix of $\epsilon^{\alpha\beta}$, one concludes that consistency is
only achieved if $\eta^{\alpha\beta}=i\lambda\epsilon^{\alpha\beta}$
for some $\lambda$. In fact, from eqs.~(\ref{eq1}),
all the $\lambda_i$ are equal to $\lambda$. In other words, there can be
just one sector. Furthermore, the commutator of every two
observables can also be worked out using eqs.~(\ref{eq4}); it follows
that $[A,B]=i\lambda(A,B)$, for arbitrary $A,B$.
There are only two possibilities: first
that $\lambda$ is non vanishing. In this case,
we end up with the usual purely quantum dynamics. Second, if
$\lambda$ vanishes, all variables are commuting.  Moreover, since the
bracket is completely determined, it coincides with the Poisson
bracket. That is, the dynamics is purely classical.

Note that this result is consistent with that found regarding
eq.~(\ref{eq3}), namely, the canonical
evolution generated by an arbitrary quadratic
Hamiltonian $\phi^\alpha\phi^\beta$ does not preserve the
semiquantized commutation relations~(\ref{eq1}).

The previous result means that there are no quotient
algebras of ${\cal A}_L$ of the form~(\ref{eq4}), and supporting a
canonical structure, which mix the classical and the quantum cases.
In passing, it can be proven \cite{Ca95}, that the bracket defined in
${\cal A}_L$ by using only linearity, antisymmetry, Leibniz
rule and the canonical relations~(\ref{eq6}),
satisfies Jacobi's identity as a byproduct, and thus this will be true
as well for any of the quotient algebras considered here if and only if
the bracket preserves the characteristic identities of that quotient algebra.

4. We conclude that, assumptions (i-iii) prevent the classical sector from
inheriting quantum fluctuations and further, assumptions (i-vi) actually
discard any non-trivial semiquantized theory.
Note that further details on
how to actually extract physical information from the observables
(e.g., expectations values in the quantum case) are
not required to reach the previous conclusion.
We comment that the approach in Ref.~\cite{Sa94}, based directly on time
ordered vacuum expectation values, is also flawed since it
breaks physical positivity of the expectation values.
It is entirely possible that there is no non-trivial
(or at least elegant) semiquantization scheme,
since, after all, such a concept is not presently
known to be physically required.

The author thanks J. Caro and C. Garc\'{\i}a-Recio for useful
criticism and for providing the proof in Ref.~\cite{Ca95}, and also
J. Kruytzer for calling my attention to Refs.~\cite{Di95a,Di95,An95a}.
This work was partially supported by the Spanish DGICYT under contract
PB92-0927 and the Junta de Andaluc\'{\i}a.

\end{document}